\documentclass[11pt,twoside]{article}

\usepackage{asp2014}

\aspSuppressVolSlug
\resetcounters

\begin{document}

\title{Astronomical Image Processing at Scale With Pegasus and Montage}

\author{G.~Bruce~Berriman$^1$, John~C.~Good $^2$, Ewa~Deelman$^3$, Ryan ~Tanaka$^3$, and Karan~Vahi$^3$  }
\affil{$^1$Caltech/IPAC-NExScI, Pasadena, CA, 91125; \email{gbb@ipac.caltech.edu}}
\affil{$^2$Caltech/IPAC-NExScI, Pasadena, CA, 91125}
\affil{$^3$USC Information Sciences Institute, Marina del Rey, CA, 90292}

\paperauthor{G.~Bruce Berriman}{gbb@ipac.caltech.edu}{0000-0001-8388-534X}{Caltech}{IPAC/NExScI}
{Pasadena}{CA}{91125}{USA}

\paperauthor{John~C.~Good}{gbb@ipac.caltech.edu}{ }{Caltech}{IPAC/NExScI}
{Pasadena}{CA}{91125}{USA}

\paperauthor{Ewa~Deelman}{deelman@isi.edu}{0000-0001-5106-503X}{USC}{Information Sciences Institute}{Marina del Rey}{CA}{90202}{USA}

\paperauthor{Ryan~Tanaka}{tanaka@isi.edu}{0000-0002-4855-6118}{USC}{Information Sciences Institute}{Marina del Rey}{CA}{90202}{USA}

\paperauthor{Karan~Vahi}{vahi@isi.edu}{0000-0001-8622-2082}{USC}{Information Sciences Institute}{Marina del Rey}{CA}{90202}{USA}

\paperauthor{Sample~Author2}{Author2Email@email.edu}{ORCID_Or_Blank}{Author2 Institution}{Author2 Department}{City}{State/Province}{Postal Code}{Country}
\paperauthor{Sample~Author3}{Author3Email@email.edu}{ORCID_Or_Blank}{Author3 Institution}{Author3 Department}{City}{State/Province}{Postal Code}{Country}



\begin{abstract}

Image processing at scale is a powerful tool for creating new data sets and integrating them with existing data sets and performing analysis and quality assurance investigations. Workflow managers offer advantages in this type of processing, which involves multiple data access and processing steps. Generally, they enable automation of the workflow by locating data and resources, recovery from failures, and monitoring of performance. In this focus demo we demonstrate how the Pegasus Workflow Manager Python API manages image processing to create mosaics with the Montage Image Mosaic engine. Since 2001, Pegasus has been developed and maintained at USC/ISI. Montage was in fact one of the first applications used to design Pegasus and optimize its performance. Pegasus has since found application in many areas of science. LIGO exploited it in making discoveries of black holes. The Vera C. Rubin Observatory used it to compare the cost and performance of processing images on cloud platforms. While these are examples of projects at large scale,  small team investigations on local clusters of machines can benefit from Pegasus as well. 

\end{abstract}



\section{Introduction}
We aim to show how astronomers can take advantage of open source tools to build workflows that can be executed on-premises machines/clusters and then scale {\it as far as possible without modification on distributed platforms}. The demo was a proof-of-concept image mosaic workflow run on a local laptop; the same workflow was run without any modifications or optimizations on the Open Science Grid (OSG) \footnote{\url{https://opensciencegrid.org/}} high-throughput platform. 

The tools we have chosen for this demo are the Pegasus Workflow Manager \footnote{\url{https://pegasus.isi.edu}} and the Montage Image Mosaic Engine \footnote{\url{http://montage.ipac.caltech.edu}}. Both are open source, operate by design on multiple platforms, support a Python API, are (reasonably) easy to use, and operate well with one another. The demo itself will create a 1$^{\circ}$ x 1 $^{\circ}$ mosaic of M17 in a Jupyter Notebook. 

\section{Pegasus and Montage}

Pegasus is a fully featured workflow management system \citep{deelman2020jocs} that enables scientists to develop portable scientific workflows, structured as directed acyclic graphs through  Python, Java, or R APIs. These workflows can then be run on heterogeneous resources, including local machines, clouds, HPC systems, and campus clusters. Some important features of Pegasus are:  monitoring and debugging capabilities via command-line interface (CLI) tools or a web dashboard; automatic handling of data movement between jobs (e.g., pulling input data from an AWS S3 bucket or sending one job’s output to another job); and running workflows with data integrity and fault tolerance in mind.

When working with Pegasus, scientists run a set of computations whose workflow can, as a starting point, be sketched on paper. The abstract workflow can be described programmatically using one of the APIs, in our case case Python. Once that is done,  the workflow will be serialized to a YAML representation, which will be consumed by Pegasus and compiled into an executable workflow intended for a specific execution environment, such as a cloud platform or a campus cluster. HTCondor sits underneath Pegasus and is responsible for running the jobs.

Montage is open source toolkit for assembling FITS images into custom mosaics \citep{2017PASP..129e8006B}.
It is written in C for performance and portability. Python binary extensions provide compiled performance on parallel platforms. The modular design of Montage, with independent modules for each step in the mosaic process, allows Montage to play well with workflow managers and enables embarrassingly parallel processing. 

\section{Try It Out}
We have created a Docker image with everything needed to run the Montage workflow just described in a Jupyter Notebook \footnote{  \url{https://github.com/pegasus-isi/ADASS21-montage-docker-image}}. This Docker image contains an installation of Pegasus and HTCondor, the Montage binaries, and the workflow notebook. You can interact with this container through Jupyter Notebooks via the browser. This container environment supports testing different workflow configurations, and contains Pegasus tutorial notebooks. 

\section{The Demo Notebook}

The notebook demonstrates proof-of-concept parallelization through creation of a small mosaic. The demo was run on a desktop machine for convenience. Because of the overhead imposed on the process by HTCondor, the real power of the approach will lie in its use on high-performance platforms.

The process involves populating three structures, maintained by Pegasus as SQLite databases: 

\begin{itemize}

\item The "replica catalog:" The input files for the processing and how to access them. 
\item The "transformations:" The Montage modules and where to run them.
\item The "workflow" enumerates the specific steps in the processing, including transforms, argument lists, which files are input and which are output, etc.  Pegasus deduces when jobs can be run in large part by when the precursor files are available.

\end{itemize}

Then the Notebook asks Pegasus to devise a processing plan, and, if successful, to submit that plan to HTCondor. Figure \ref{fig1} shows a screenshot of one step in the process.

\articlefigure{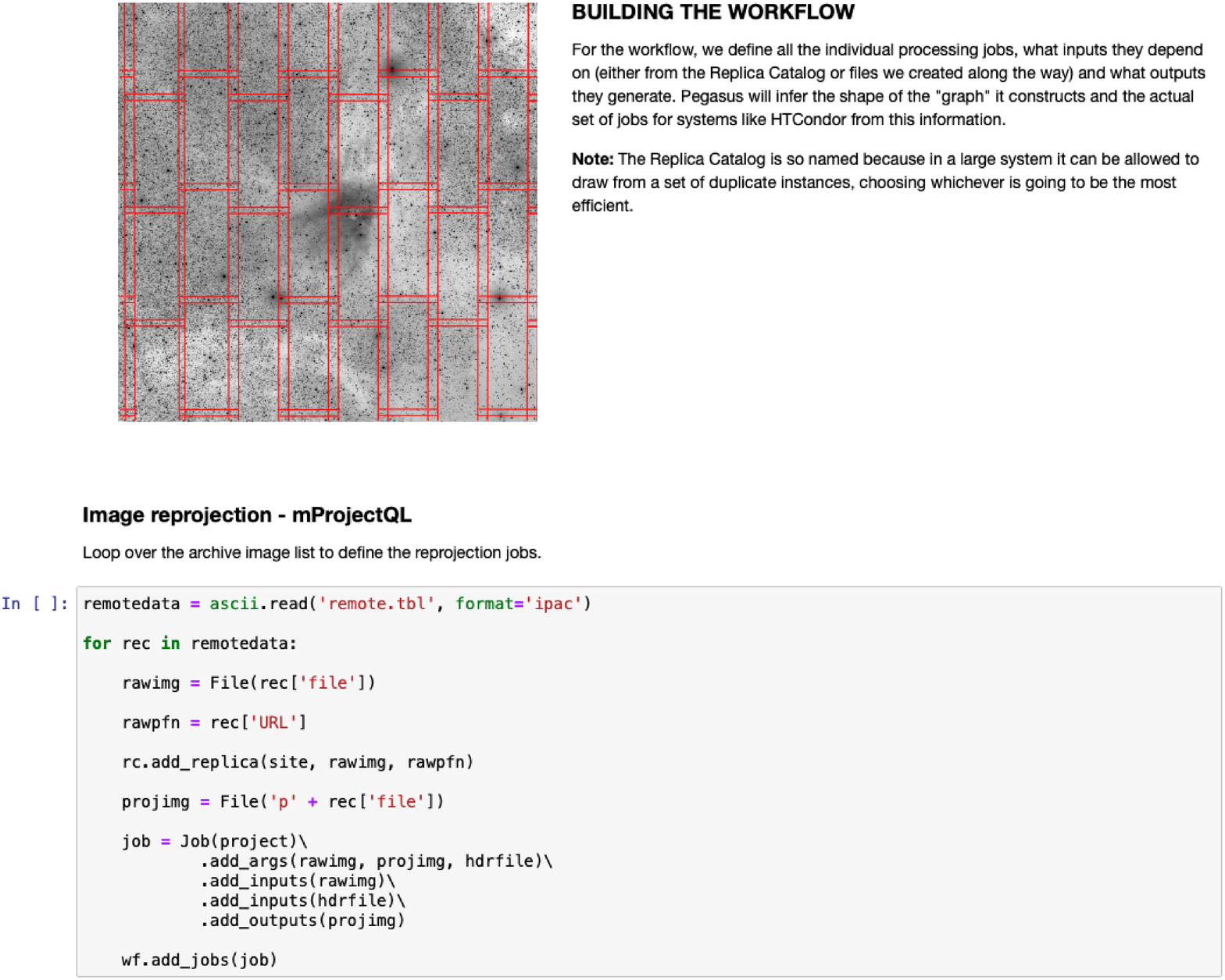}{fig1}{Screenshot of a section of the Notebook for building the Pegasus Montage workflow. }

\section{Running the Workflow on a Parallel Platform: The  Open Science Grid}  

We ran the Montage workflow on the OSG \citep{osg09} with minimal modifications to the Python workflow script to  utilize OSG-specific features: 
\begin{itemize}

\item Data movements were managed by OSG’s StashCache data infrastructure, which caches files in an opportunistic manner.
\item The workflow used RedHat Enterprise 7 Singularity containers, which are cached within OSG’s CernVM-File system (CVMFS).  Setup required changing only a few lines in our workflow generation script. 

\end{itemize}

Table 1 summarized metrics Pegasus reported on the 332 jobs (including jobs added by Pegasus to handle data staging and cleanup).

\begin{table}[!ht]
\caption{Pegasus Montage Workflow Metrics on OSG}
\smallskip
\begin{center}
{\small
\begin{tabular}{ll}
\tableline
\noalign{\smallskip}
Metric  & Value \\
\tableline
\noalign{\smallskip}
Number of Jobs & 332 \\
Workflow Wall Time  & 54m 43s \\
Cumulative Job Wall Time & 36m 48s \\
Cumulative Job Wall Time (from submit side)&  2h 38m \\
Integrity Checking Time Spent & 3.57s (for 357 files) 

\end{tabular}
}
\end{center}
\end{table}

\noindent 

The wall times reveal considerable queuing overhead associated with running the workflows on OSG. Because the overheads are incurred at the time of job submission, it is beneficial to cluster jobs together to pay a single overhead for a "cluster" of jobs. Pegasus can do this clustering automatically based on user-specified criteria. Pegasus also adds overheads such as integrity checking, but these, by contrast, are negligible and in this case took only  3.57 seconds.

\section{Conclusions}

We have successfully run Pegasus and Montage on a laptop and on the OSG and we have made our setup available to others for replication and reuse. Our next steps include optimizing performance on parallel platforms through clustering and running the workflow on commercial cloud platforms.

\acknowledgements Pegasus is funded by the National Science Foundation under OAC SI2-SSI program grant 1664162. Previously, NSF funded Pegasus under OCI SDCI program grant 0722019 and OCI SI2-SSI program grant 1148515. Montage is funded by the National Science Foundation grant awards 1835379, 1642453 and 1440620. OSG is supported by the National Science Foundation grant award 2030508.

\bibliography{D0-001}  

\begin{thebibliography}{}
\expandafter\ifx\csname natexlab\endcsname\relax\def\natexlab#1{#1}\fi
\expandafter\ifx\csname url\endcsname\relax
  \def\url#1{\texttt{#1}}\fi
\expandafter\ifx\csname urlprefix\endcsname\relax\def\urlprefix{URL }\fi
\providecommand{\eprint}[2][]{\url{#2}}

\bibitem[{{Berriman} \& {Good}(2017)}]{2017PASP..129e8006B}
{Berriman}, G.~B., \& {Good}, J.~C. 2017, \pasp, 129, 058006.
  \eprint{1702.02593}

\bibitem[{Deelman et~al.(2021)Deelman, Ferreira~da Silva, Vahi, Rynge, Mayani,
  Tanaka, Whitcup, \& Livny}]{deelman2020jocs}
Deelman, E., Ferreira~da Silva, R., Vahi, K., Rynge, M., Mayani, R., Tanaka,
  R., Whitcup, W., \& Livny, M. 2021, Journal of Computational Science, 52,
  101200

\bibitem[{Sfiligoi et~al.(2009)Sfiligoi, Bradley, Holzman, Mhashilkar, Padhi,
  \& Wurthwein}]{osg09}
Sfiligoi, I., Bradley, D.~C., Holzman, B., Mhashilkar, P., Padhi, S., \&
  Wurthwein, F. 2009, in 2009 WRI World Congress on Computer Science and
  Information Engineering, vol.~2 of 2, 428

\end{thebibliography}


\end{document}